\documentclass[aps,pre,twocolumn,showpacs,amssymb]{revtex4}
\usepackage[dvips]{graphicx}
\usepackage{dcolumn}% Align table columns on decimal point
\usepackage{bm}% bold math
\def\gs{\gtrsim}
\def\ls{\lesssim}
\def\be{\begin{equation}}
\def\en{\end{equation}}    
\def\gs{\gtrsim}
\def\ls{\lesssim}
\newcommand{\bi}[1]{\mbox{\boldmath$#1$}}

\def\p{\partial}
\def\bea{\begin{eqnarray}}
\def\ena{\end{eqnarray}}
\newcommand{\ppp}[3]{{\bigg(}\frac{\partial {#1}}{\partial {#2}}{\bigg )}_{#3}}

\newcommand{\ppd}[3]{{\bigg(}\frac{\delta {#1}}{\delta {#2}}{\bigg )}_{#3}}

\begin{document}
\draft
\bibliographystyle{prsty}
\title{Spreading with evaporation and condensation in one-component fluids}
%\title{Spreading and evaporation of droplets }
\author{Ryohei  Teshigawara  and Akira Onuki}
\address{Department of Physics, Kyoto University, Kyoto 606-8502}
\date{\today}

\begin{abstract} 
We investigate the dynamics of spreading of 
a small liquid droplet in gas 
 in a one-component simple 
  fluid, where the temperature is inhomogeneous 
around $0.9T_c$  and latent heat is  released or generated at the interface 
  upon evaporation or   condensation (with 
  $T_c$ being the critical temperature). 
   In the scheme of 
 the  dynamic van der Waals theory, 
   the hydrodynamic equations 
containing the gradient stress  are solved 
in the axisymmetric geometry. We assume that the  
 substrate has  a finite thickness  
and  its temperature obeys 
 the thermal diffusion  equation. 
A precursor film  then spreads ahead of 
the bulk droplet itself  
in the complete wetting condition. 
Cooling the substrate enhances  
condensation of gas onto the advancing film, which   
mostly takes  place near the film edge and can be the 
dominant mechanism of the film growth in a late stage. 
The generated  latent heat produces  a  
temperature peak or a hot spot  
in the gas  region near the film edge.   
On the other hand, 
heating  the substrate induces evaporation 
all over the interface. 
For weak  heating, 
 a steady-state  circular thin 
film can be  formed on the substrate. 
For stronger heating, 
 evaporation dominates over condensation, 
leading to eventual disappearance of the liquid region. 
\end{abstract}

\pacs{68.03.Fg, 68.08.Bc, 44.35.+c,  64.70.F-}

\maketitle

\pagestyle{empty}

%%%%%%  introduction%%%%%%%%%%
\section{Introduction}
\setcounter{equation}{0}

Extensive efforts 
have been made on the static and dynamic properties of 
wetting transitions  for various fluids and substrates 
both theoretically and experimentally 
 \cite{PG}. 
In particular,  spreading of a liquid 
has been studied by many groups 
\cite{PG,Hardy,Dussan,Joanny,Bonn}, since it 
  is of great importance in a number of  
practical situations such
as lubrication, adhesion, and painting. 
Hydrodynamic theories were developed 
for  spreading of an involatile liquid droplet in gas  
in an early stage of the theoretical research 
\cite{PG,Joanny}. 
A unique feature revealed by experiments 
 \cite{Hardy,Leger,Cazabat} is  that a thin 
 precursor film is formed ahead of the liquid droplet 
 itself in the complete wetting condition.  
Hardy first reported its formation  
ascribing its origin to condensation at the film 
edge \cite{Hardy}, but it has been observed also 
for involatile fluids \cite{Leger,Cazabat}. 
To understand nanometer-scale  spreading processes,  
a number of microscopic  simulations have been performed 
mainly for fluids composed of chain-like molecules 
\cite{Koplik,Ni,Monte,p1,p2,p3,Binder,Grest}.

However, understanding of 
the wetting dynamics of volatile liquids is still 
inadequate. We mention some examples 
where evaporation and condensation come into play. 
In their molecular dynamic simulation \cite{p3},  
Koplik {\it et al.} observed 
evaporation of a droplet and a decrease 
of the contact angle upon heating a substrate 
in the partial wetting condition. 
In their experiment \cite{Caza}, 
Gu$\acute{\rm e}$na {\it et al.} observed that 
a weakly volatile droplet 
spread  as  an  involatile droplet   in an initial stage 
but   disappeared  after 
a long time due to evaporation 
in the complete  wetting condition. 
In a near-critical one-component fluid \cite{Hegseth}, 
{Hegseth} {\it et al.} 
observed that  a bubble was 
attracted to a heated wall even when it was completely wetted
by liquid in equilibrium (at zero heat flux), where  the apparent 
contact angle of a bubble 
increased  with the heat flux.

In addition to  spreading 
on a heated or cooled substrate, 
there are a variety of situations such as 
droplet evaporation \cite{Nature,Larson,Bonn-e,But,Teshi}, 
boiling  on a heated substrate \cite{Str,Nikolayev,Stephan}, 
and  motion of a bubble  
suspended in liquid  \cite{Beysens,Kanatani}, where   
latent heat  generated or released at the interface 
drastically influences the hydrodynamic processes. 
In particular, a large temperature gradient and a large heat flux 
 should be produced  around  
 the edge of a liquid film  
 or the contact line of a droplet or  bubble 
 on a substrate \cite{Nikolayev,Teshi}. 
The   temperature and velocity profiles 
should be    highly singular in these narrow regions.   
Here  an experiment by  
H$\ddot{\rm o}$hmann and Stephan 
\cite{Stephan} is noteworthy. 
They  observed a sharp  drop in the substrate temperature  
near the contact line of a growing bubble in boiling. 
%However, this effect  
%has not yet been well studied in the literature. 
Furthermore, we should stress  relevance of 
the Marangoni flow in multi-component fluids 
in two-phase hydrodynamics \cite{Str,Larson,maran}, 
where temperature and concentration  variations  
cause a  surface tension gradient and a balance 
of the tangential stress  
induces   a flow  on the droplet scale. 
%However, simulations of 
%the Marangoni effect have 
%rarely been performed \cite{Y}. 

In hydrodynamic theories, 
the gas-liquid transition has been  
included  with the aid of 
a phenomenological input of  the  
 evaporation rate on the interface $J$. 
Some authors   \cite{Nature,Larson,Bonn-e} assumed 
  the form  $J(r,t) =J_0/\sqrt{r_e(t)^2 -r^2}$ for a thin 
circular  droplet   as a function of 
the distance $r$ from the droplet center, where 
 $r_e(t)$ is the film radius and  $J_0$ is a constant. 
In the framework of the lubrication  theory, 
Anderson and Dabis \cite{Davis} examined  
spreading of a thin volatile 
droplet on a heated substrate 
by assuming  the form  $J= (T_I-T_{\rm cx})/K^*$, 
where $T_{I}$ is  the interface temperature, 
$ T_{\rm cx}$ is  the saturation (coexistence) temperature, and  
  $K^*$ is  a kinetic coefficient. In these papers,  
the  dynamical processes in the  gas  
have   been neglected.

Various mesoscopic (coarse-grained) simulation methods 
have also been used to investigate two-fluid 
hydrodynamics, where the interface has a  finite thickness. 
 We mention  phase field models 
of fluids (mostly treating 
incompressible binary mixtures)  
\cite{Se,Jq,In,Ta,Araki,Pooly,Jasnow,Jamet,Borcia,OnukiPRL,OnukiV,phase1,G,Y,Palmer,Ohashi},  where   the gradient stress is included  
in the hydrodynamic equations 
(see a review in Ref.\cite{phase1}).  
In particular, some authors  numerically  studied liquid-liquid phase 
separation  in heat flow \cite{Jasnow,Araki,Pooly,G}, but 
 these authors treated symmetric binary mixtures  
without latent  heat.  
Recently, one of the present authors 
developed a phase field model for compressible fluids 
with inhomogeneous 
temperature, which is called  
the  dynamic van der Waals model  
\cite{OnukiPRL,OnukiV}. In its framework, 
 we  may  describe   
 gas-liquid transitions  and  
convective latent heat transport 
 without  assuming  any   evaporation formula. 
In one of its applications  \cite{Teshi}, 
it was used to investigate 
 evaporation of an axisymmetric  droplet on 
 a heated substrate in a one-component system. 
Our finding  there is  that  evaporation  occurs 
 mostly near   the contact line. 
We also mention the lattice Boltzmann method 
to simulate the continuum equations, where 
the molecular velocity takes discrete values   
\cite{Palmer,Y,Pooly,In,Ohashi}. 
However, this method  has not yet been fully  developed   
 to describe evaporation and condensation.

In  this  paper, we will  
 simulate  spreading      
using the  dynamic van der Waals model 
 \cite{OnukiPRL,OnukiV}. 
%In this scheme the temperature changes 
%smoothly across the interface during evaporation.  
We  will treat a one-component fluid  
in a temperature range 
around  $0.9T_c$, where 
the gas  and  liquid densities  
are not much separated. 
Namely,  we will approach the 
problem relatively close to the critical point. 
Then the mean free path in the gas  
is not long,  so that 
 the temerature may be treated to be 
continuous across an interface  
in nonequilibrium. 
When the gas  is dilute,  
 the phase field 
aproach becomes  more difficult to  treat  gas flow 
produced by evaporation and condenssation. 
It is known that     the temperature 
near  an interface changes sharply in the gas  
over  the mean free path  
during evaporation  \cite{Ward}.

The organization of this paper is as follows. 
 In  Sec.II, we  will present the dynamic  equations  with 
appropriate boundary conditions.  
In  Sec.III, the  simulation method will be explained.  
In Sec.IV,    numerical results  
of spreading  will be given 
for cooling and heating the substrate.

%%%%%%%%   dynamic van der Waals theory     %%%%%%%%%%
%\section{Theoretical background} 

\section{Dynamic van der Waals theory}
\setcounter{equation}{0}

%In the dynamic van der Waals theory of 
%the fundamental dynamical variables are  $n=n({\bi r},t)$, 
% $e=e({\bi r},t)$, 
%and the velocity field ${\bi v}= {\bi v}({\bi r},t)$. 
When we discuss phase transitions 
with  inhomogeneous temperature, 
the free energy functional 
 is not  well defined.  In such  cases, we  should 
start with an entropy functional including a gradient 
contribution,  which is  determined by the 
number density $n=n({\bi r},t)$ and the 
internal  energy density $e=e({\bi r},t)$ 
in one-component fluids. 
Here we present  minimal forms  of 
the entropy functional  and 
the  dynamic  equations needed for our simulation.

\subsection{Entropy formalism}

We introduce  a  local entropy density 
$\hat{S}=\hat{S}({\bi r},t)$   consisting   of  regular 
 and  gradient terms as \cite{OnukiPRL,OnukiV} 
\be 
\hat{S}= ns(n,e)-
\frac{1}{2}C | \nabla n |^2.
\en
Here $s=s({\bi r},t)$ is the  entropy per particle  depending   
on  $n$ and $e$. The coefficient  $C$ of the gradient 
term can  depend on $n$ \cite{OnukiV}, 
but it  will be assumed to be a positive constant 
independent of $n$. The gradient entropy is  
negative and is particularly important  in the 
interface region. The  entropy functional 
is the space integral 
${\cal S}_b\equiv \int d{\bi r}\hat{S}$  
 in the bulk region. 
As a function of 
$n$ and $e$, the temperature $T$ is determined from 
\be 
\frac{1}{T}= \ppd{{\cal S}_b}{e}{n}= 
n\ppp{s}{e}{n}. 
\en  
The generalized chemical potential $\hat{\mu}$ 
including the gradient part is of the form, 
\be 
\hat{\mu} = -T \ppd{{\cal S}_b}{n}{e}= \mu -TC \nabla^2n, 
\en 
where $\mu=-T [\p (ns)/\p n]_e$ is the usual chemical potential per particle. 
In equilibrium $T$ and $\hat\mu$ are homogeneous constants. 
In this paper,  we introduce the gradient entropy as in Eq.(2.1), 
 neglecting  the gradient energy  
\cite{OnukiPRL,OnukiV}.  Then the total internal   energy in the bulk  is 
simply  the integral $\int d{\bi r} e$.

In the van  der Waals theory \cite{Onukibook}, 
 fluids are characterized by 
the molecular volume $v_0$ and  the 
 pair-interaction energy $\epsilon$. 
As a function of $n$ and $e$,   $s$ is written as 
\be 
 s =  k_{B}      \ln [(e/n+ \epsilon v_0n)^{3/2}
  ({1}/{v_0n}-1 )]  +s_0, 
\en 
where $s_0/k_B= \ln [v_0 ({m/3\pi\hbar^2})^{3/2}]+5/2$ 
with $m$ being the molecular mass. 
We define  $T$ as in Eq.(2.2) 
to obtain the well-known expression for the internal energy   
$
e=3nk_BT/2-\epsilon v_0n^2   
$ and the pressure  
\be 
p= n\mu+TS-e= nk_BT/(1-v_0n)- \epsilon v_0n^2.
\en 
The  critical density, temperature, 
and pressure read 
\be \
n_c =1/3v_0,\quad 
T_c=8\epsilon/27k_B, \quad 
p_c= \epsilon/27v_0, 
\en 
respectively. 
Macroscopic gas-liquid  coexistence with a planar interface  
is realized for $T<T_c$ and   
at  the saturated vapor pressure $p=p_{cx}(T)$. 
With introduction of the gradient entropy, 
there arises a length  $\ell$ defined by  
\be 
\ell =(C/2k_B v_0)^{1/2},  
\en 
in addition to the molecular diameter $\sim v_0^{1/3}$.  
From Eq.(2.3) the  correlation length $\xi$ is 
defined by $\xi^{-2}= (\p \mu/\p n)_T /TC$, so $\xi$ is  
proportional to $\ell$ as 
\be 
\xi/ \ell=  n (2v_0k_BTK_T)^{1/2}, 
\en 
where $K_T= (\p n/\p p)_T/n$ is the isothermal compressibility. 
The interface 
thickness is of order $\xi$ in two-phase coexistence. 
%and  $\gamma$ is  of order 
%$k_BT(1-T/T_c)^{3/2} \ell/v_0$  
%in our  mean-field theory. 
The ratio $\ell/v_0^{1/3}$
should be of  order unity for real  simple fluids. 
However, we   may treat   
$\ell$ as  an arbitrary parameter 
in our  phase field scheme. 

\subsection{Hydrodynamic equations}

We set up the  hydrodynamic equations 
from the principle of 
positive entropy production  in 
nonequilibrium \cite{Landau}. 
The mass density $\rho=mn$ obeys the continuity equation,  
\be 
\frac{\p}{\p t} \rho = - \nabla \cdot(\rho{\bi v}),
\en 
where $\bi v$ is 
the velocity field assumed to vanish 
on all the boundaries. 
In the presence of an externally applied  potential 
field $U({\bi r})$ (per unit mass), 
we write the equation for the 
momentum density $\rho {\bi v}$ as  
\be 
\frac{\p}{\p t}\rho {\bi v}
=-\nabla \cdot (\rho  {\bi v}{\bi v}+
 \tensor{\Pi}-\tensor{\sigma} ) -\rho \nabla U.
\en
In our previous work\cite{OnukiV} we set 
$U= gz$ for a gravitational field with $g$ being 
the gravity acceleration. We note  that 
$U$ may also represent  the van der Waals interaction 
 between the fluid particles and the solid 
 depending the distance from the wall \cite{PG}. 
The stress tensor is divided into three parts. 
The $\rho{\bi v}{\bi v}$ is the inertial part. The 
 $\tensor{\Pi}=\{\Pi_{ij}\}$ is the reversible part   
including   the gradient stress tensor,
\bea
\Pi_{ij} &=& \bigg [
p  - CT( n\nabla^2 n+\frac{1}{2} |\nabla n|^2)\bigg]\delta_{ij}
\nonumber\\
&&  +CT (\nabla_i n)(\nabla_j n), 
\ena 
where  ${ p}$ is the van der Waals pressure 
in Eq.(2.5).  Hereafter $\nabla_i= \p /\p x_i$ 
with $x_i$ representing $x$, $y$, or  $z$.   
The  $\tensor{\sigma}= \{{\sigma}_{ij}\}$ 
is the  viscous stress tensor  expressed as 
\be 
{\sigma}_{ij}
=\eta(\nabla_i v_j+\nabla_j v_i)  +
(\zeta-2\eta/3) (\nabla \cdot {\bi v})\delta_{ij} ,
\en 
 in terms of the 
shear viscosity $\eta$ and the bulk viscosity 
$\zeta$. Including the kinetic energy density 
and the potential energy,   we define the (total) energy density by  
$  
e_{\rm T}=\hat{e}+ \rho {\bi v}^2/2+\rho U. 
$ 
It   is a conserved quantity governed by \cite{gravity} 
\be 
\frac{\p}{\p t} {e}_{\rm T}=
 -\nabla\cdot\bigg[ e_{\rm T}{\bi v} + 
(\tensor{\Pi}-\tensor{\sigma})\cdot{\bi v} 
-\lambda \nabla T\bigg],
\en 
where $\lambda$ is 
the thermal conductivity. 
With these hydrodynamic equations including the 
gradient contributions, 
the entropy density  $\hat S$ 
in Eq.(1) obeys 
\be\label{eq:entropy_dev}
\frac{\p\hat{S}}{\p t}+\nabla\cdot \bigg[
\hat{S}{\bi v}-Cn(\nabla\cdot {\bi v})\nabla n
-\frac{\lambda}{T}\nabla T
\bigg]= 
\frac{\dot{\epsilon}_v+\dot{\epsilon}_\theta}{T}, 
\en
where the right hand side is  
the nonnegative-definite entropy production rate with 
\be   
\dot{\epsilon}_v= \sum_{ij}\sigma_{ij}\nabla_j v_i,
\quad 
\dot{\epsilon_\theta}= \lambda(\nabla T)^2/T.
\en  
In passing,  the constant  $s_0$ in Eq.(2.4) may be omitted  
in Eq.(2.14) owing to the continuity equation (2.9). 

\subsection{Boundary conditions}
We  assume 
the no-slip boundary condition,  
\be 
 {\bi v}={\bi 0},
\en  
on all the  boundaries for simplicity. 
However,  a number of  molecular dynamic simulations 
have shown   that  a  slip 
of the tangential fluid velocity   becomes significant  
around a  moving contact line 
\cite{slip,Qian}.

We assume  the 
surface entropy  density $\sigma_s(n_s)$ 
and the surface energy 
density  $\epsilon_s(n_s)$ 
depending  on the fluid density 
 at the surface, written as $n_s$.  
The total entropy 
including the surface contribution 
is of the form, 
\be 
{\cal S}_{\rm tot}= \int d{\bi r} \hat{S} + \int da \sigma_s,   
\en 
where $\int da$ is  the surface integral 
over the boundaries. The total fluid energy is given by 
\be 
{\cal E}_{\rm tot}= \int d{\bi r}(e+\frac{1}{2}\rho
{\bi v}^2 + \rho U) 
+ \int da e_s.
\en 
We assume that there is  no 
strong adsorption of the fluid particles onto 
the boundary walls. 
 The fluid density is 
 continuously connected from the bulk  to the boundary 
 surfaces; for example,   we have 
 $n_s(x,y,t) =\lim_{z\to +0}n({\bi r},t)$ at $z=0$.  
 Then the  total particle 
number of the fluid in the cell 
is  the bulk integral ${\cal N}=\int d{\bi r} n$.

We  assume that 
the temperatures in the fluid and in the solid are 
continuously connected at the surfaces. 
The  temperature on  the substrate 
is then well-defined 
and we  may  introduce the surface  
Helmholtz free energy density,  
\be 
f_s=e_s -T \sigma_s.
\en 
As  the surface boundary condition, we require 
\be 
TC\hat{\bi \nu}_b\cdot {\nabla n}= - \ppp{f_s}{n_s}{T},  
\en 
where $\hat{\bi \nu}_b$ is the outward surface normal unit vector. 
This   boundary condition  has been 
obtained in equilibrium with homogeneous $T$ by 
 minimization of the total Helmholtz (Ginzburg-Landau) free energy,
\be 
F_{\rm tot}= \int d{\bi r}(e- T\hat{S}) + 
\int da f_s.
\en    
We assume this  boundary condition in Eq.(2.20) 
even in nonequilibrium. 
Then use of Eq.(2.14) yields \cite{Landau} 
\be 
\frac{d}{dt} {\cal S}_{\rm tot}
= \int d{\bi r}\frac{\dot{\epsilon}_v+\dot{\epsilon}_\theta}{T} 
+\int da \frac{\hat{\bi \nu}_b\cdot\lambda\nabla T 
+{\dot{e}}_{s} }{T}.
\en 
where $\dot{e}_{s}=\p {e}_{s}/\p t= (\p e_s/\p n_s)(\p n_s/\p t)$. 
The first  term in the right hand side is the 
bulk entropy production rate, while the second term is the 
the surface integral of the 
heat flux from the solid divided by $T$ 
or the entropy input from the solid to the fluid.

In this paper, 
we present simulation results with $U=0$ 
for simplicity.  In our previous work \cite{OnukiV} 
 a large gravity field was assumed in boiling. 
In future we should investigate  the effect of the long-range 
van der Waals interaction in  the wetting dynamics.

\section{Simulation Method}
\setcounter{equation}{0}

In our phase field simulation, we integrated  
the continuity equation (2.9), 
the  momentum equation (2.10), and  
the entropy equation (2.14),  
not using   the  energy equation (2.13), as 
in our previous simulation \cite{Teshi}.  
With this method, if there is no  applied heat flow, 
temperature and velocity   gradients 
tend  to vanish at long times 
in the whole space including the interface 
region.  This numerical stability is achieved  
because the  heat production 
 rate  $\dot{\epsilon}_v+\dot{\epsilon}_\theta \ge 0$ 
 appears explicitly in the entropy equation,  so that 
${d}{\cal S}_{\rm tot}/dt \ge 0$ in Eq.(2.22) 
without applied heat flow.
We can thus  successfully describe 
the temperature and velocity  near the film edge 
(those  around the contact line of an evaporating 
droplet in Ref.\cite{Teshi}).

It is worth noting that  many authors  
have encountered   a parasitic flow 
 around a curved interface 
in numerically solving the hydrodynamic equations 
  in two-phase states \cite{para,Ohashi}.  
It remains nonvanishing 
 even when the system should tend to 
 equilibrium without  applied heat flow. 
 It is an  artificial flow, 
  since its magnitude depends 
on the discretization method.

\subsection{Fluid  in a cylindrical cell}

We suppose a cylindrical cell. 
Our model fluid is in the region 
$0\leq z\leq H$ and $0\leq r=({x^2+y^2})^{1/2} \leq L$, 
where  $H= 300\Delta x$ and $L=400\Delta x$ 
with  $\Delta x$ being  the simulation mesh length. 
 The velocity field $\bi v$ vanishes 
on all the boundaries.  In this axisymmetric geometry, 
 all the  variables are assumed to  depend 
only on $z,$ $r$ and $t$.  The integration 
of the dynamic equations is  
on a $200\times 400$ lattice in the fluid region. 
 We  set $\Delta x=\ell/2$, where  $\ell$ 
 is defined  in Eq.(2.7).  We will 
 measure space in units of $\ell$.  
  Then $H=150$ and $L=200$ in units of $\ell$.

The transport coefficients are 
 proportional to $n$ as   
\be 
\eta= \zeta= \nu_0 mn, \quad \lambda=k_B\nu_0 n.
\en  
These   coefficients 
 are larger in liquid than  in gas by the density ratio 
$n_\ell/n_g (\sim 5$ in our simulation). 
The  kinematic viscosity 
$\nu_0=\eta/mn$ is a constant. 
We will  measure time in units of the viscous relaxation time,  
\be 
\tau_0=\ell^2/\nu_0= C/2k_Bv_0\nu_0,
\en 
on the scale of $\ell$.  
We will measure velocities 
in units of $\ell/\tau_0=\nu_0/\ell$. 
The time  mesh size of our simulation 
 is $\Delta t= 0.01\tau_0$. 
Away  from the criticality, 
the thermal diffusivity $D_T= \lambda/ C_p$  
is of order $\nu_0$ and   and the   Prandtl number 
$Pr= \nu_0/D_T$ is of order unity,  so  
$\tau_0$ is also the thermal 
relaxation time on the scale of $\ell$.
Here the  isobaric specific heat  $C_p$  
per unit volume is of order $n$ 
far from the criticality, while  it grows 
in its vicinity.   With Eq.(3.1),  
there arises a  dimensionless number given by  
\be 
\sigma=m\nu_0^2/\epsilon\ell^2=m\ell^2/\epsilon\tau_0^2 .
\en  
The transport coefficients 
are  proportional to $\sigma^{1/2}$. 
 In this paper we set $\sigma=0.06$, for which 
sound waves are well-defined as oscillatory modes  
for wavelengths longer than 
$\ell$ (see Fig.5) \cite{OnukiV}.

The temperature at the top $z=H$ 
is fixed at $T_H$, while  the side wall 
at $r=L$ is thermally insulating or 
$ 
\hat{\bi \nu}_b\cdot \nabla T ={\p T}/{\p r}=0 
$
at $r=L$.  The  boundary condition 
of the density $n$ on 
the substrate  $z=0$ is given by    
\be 
v_0 \ell  \frac{\p n}{\p z}=-\Phi_1,
\en 
where  $\Phi_1$  arises  from the short-range 
 interaction between the fluid and the solid wall 
\cite{PG,Yeomans}.  We treat $\Phi_1$ as a parameter 
independent of $T$. From Eq.(2.19) 
this can be the case where  $e_s=0$ and $\sigma_s(n_s) = 
(\Phi_1 C/v_0\ell) n_s$.   
For example, at $T=0.875T_c$, 
the contact angle is zero at $\Phi_1\cong 0.060$ 
and the wall is completely  
wetted by liquid for larger $\Phi_1$.
Furthermore, we  set 
$ {\p n}/{\p z}=0$ on   the top plate at $z=H$ 
and $ {\p n}/{\p r}=0$ on the side wall  at $r=L$.

%\subsection{Eliminating Parasitic flow 
%by solving the entropy equation}

\subsection{Solid substrate}

In our previous work, we assumed a constant 
temperature at the bottom plate $z=0$ 
\cite{OnukiPRL,OnukiV,Teshi}. 
In this paper, we suppose 
 the presence of a sold wall  
in the region $-H_w\leq z \leq 0$ 
and $0\leq r=({x^2+y^2})^{1/2}\leq L$, where 
its  thickness  is 
$H_w= 100\Delta x=50\ell=H/3$. The temperature in the solid 
obeys the thermal diffusion equation, 
\be 
{C_w} \frac{\p T}{\p t} = {\lambda_w} \nabla^2 T ,
\en 
where  $C_w$ is the heat capacity (per unit volume) and 
$\lambda_w$ is the thermal conductivity  of the solid. 
The temperature $T(r,z,t)$ is continuous across the substrate $z=0$. 
In our simulation, the thermal diffusivity in the solid 
is given by  $D_w=\lambda_w/C_w= 
400\nu_0$, while 
the thermal diffusivity of the fluid $D_T$ 
is of order $\nu_0$ away from the criticality. 
Thus the thermal relaxation time in the substrate 
is  $H_w^2/D_w= 25\tau_0$, which is shorter 
than typical spreading times to follow. 
Because $D_w \gg D_T$,  
we integrated  Eq.(3.5) using the implicit 
Crank-Nicolson method 
on a $100\times 400$ lattice.

In this paper, the temperature $T$ 
at the substrate bottom  
$z=-H_w$ is held fixed at a  constant $T_w$. 
That is, for any $r$, we assume  
\be 
T(r,-H_w)= T_w.
\en  
Heating (cooling) of the fluid occurs 
when  $T_w$ is higher (lower) than 
the initial fluid temperature $T_0$. 
There is no heat flux through the side wall, so 
 $\p T/\p r=0$ at $r=L$ as in the fluid region. 
From the energy conservation at the 
boundary,
 the heat flux on the substrate  surface is continuous as 
\be 
({\lambda_w} {T'})_{z=-0}   = ({\lambda}  T')_{z= +0},   
\en 
where $T'= \p T/\p z$.  This holds if 
there is  no appreciable  variation of the surface energy density 
$\epsilon_s$.  We define the parameter, 
\be 
\Lambda=\lambda/( nv_0\lambda_w)= k_B \nu_0/ v_0\lambda_w.
\en 
Then $(T')_{z=-0}= \Lambda n_sv_0 (T')_{z=+0}$ on the substrate. 
In this paper, 
$\Lambda$ is set equal to  $0.002$ or $0.2$. 
We found that  the boundary 
temperature at $z=0$ 
is nearly isothermal at $T=T_w$ for $\Lambda=0.002$ 
but  considerably inhomogeneous 
around  the edge for $\Lambda=0.2$.

\subsection{Preparation of the initial state and weak  
adsorption preexisting before spreading} 

To prepare the initial state, we 
first   placed  a semispheric 
liquid droplet with radius $R=40\ell$ 
on the substrate $z=0$  
with  gas  surrounding it. Here we set  
$\Phi_1=0$ to suppress  adsorption of the fluid 
to the solid. The temperature and pressure were 
$T=T_0= 0.875T_c$ and $p= p_{cx}(T_0) =0.573p_c$ 
on the coexistence line 
in the fluid. 
The  liquid and gas densities 
were  those on the coexistence 
curve,   $n_\ell^0= 
0.579v_0^{-1}$  in liquid 
and $n_g^0= 0.123v_0^{-1}$ in gas.  
The entropy difference  between 
the two phases is $2.1k_B$ per particle.  
The total particle number is $N 
 = 2\pi (n_\ell^0-n_g^0) R^3/3+ \pi n_g^0 L^2H=
%1.64\times 10^6\ell^3/v_0$. 
1.61\times 10^6\ell^3/v_0$.
The  particle number in the droplet    
is   about $5\%$ of $N$.

Next,  we  waited for an equilibration 
 time of $10^4$ with $\Phi_1=0$.  The contact angle was  
 kept at $\pi/2$ 
and ${\hat{\bi \nu}}_b \cdot \nabla  n=0$ on all the boundary 
surfaces. However, 
the  liquid and gas pressures were 
slightly changed to $0.608p_c$ 
and $0.575p_c$, respectively. 
The pressure difference $\Delta p= 0.033 p_c$ 
is  equal to $2\gamma/R$ from  
 the Laplace law.  In accord with this, 
  the surface tension $\gamma$ at 
$T=T_0$ is given by 
$
\gamma= 0.66\ell p_c  
$ in our model.  
As a result, the  liquid density 
was  increased to  $0.583v_0^{-1}$ 
and the droplet radius was decreased to $38\ell$.  
%Phi_1(old)=\Phi_1/CT and \Phi_1= 0.0596 <--> \Phi_1(old)=0.115 (C=2)
After this equilibration 
we hereafter set $t=0$ as the origin of the time axis.

At $t=0$, we changed the wetting parameter 
$\Phi_1$ in the boundary condition (3.4) 
 from 0 to $0.0610$ to realize the complete 
wetting condition. 
Before appreciable  spreading,  
weak adsorption of the fluid has been  induced 
on the substrate  in a short time of 
order unity (in units of  $\tau_0$). 
 For small $\Phi_1$ and 
  away from  the contact line, 
this preexisting    density deviation, written as  
$\delta n(z)$,      
is of  the exponential  form, 
\be 
 \delta n(z) = ({\xi \Phi_1}/{v_0\ell}) e^{-z/\xi} ,
\en 
in terms of the correlation length $\xi$.  
Note that homogeneity of $\hat{\mu}$ 
in Eq.(2.3) yields  
 $(\xi^{-2}-\p^2/\p z^2)\delta n=0$ 
 in the linear order, leading to 
  Eq.(3.9) under  Eq.(3.4).   
The $z$ integration of  $\delta n(z)$ is   the 
excess adsorption, 
\be 
\Gamma_{\rm ad}= 
\xi^2 \Phi_1/{v_0\ell}.
\en  
In the gas at $T=0.875T_c$,   Eq.(2.8) gives 
 $\xi=1.68\ell$,  leading to  $ \Gamma_{\rm ad} =0.24 \ell/v_0$.
 We shall see that this adsorption  is one order of magnitude 
smaller than that due to a precursor film $(\sim 2.5\ell/v_0$ in Fig.6 below).

\section{Spreading on a cooled substrate}
\setcounter{equation}{0}

We present  numerical 
results of droplet spreading 
on a cooler  substrate. 
At $t=0$ the bottom temperature 
 $T_w$  at $z=-H_w$ 
 was lowered from   $T_0= 0.875T_c$ 
to $0.870T_c$ except for two curves in Fig.2 (for which 
$T_w=T_0$  even for $t>0$)). 
The top  temperature at $z=H$ 
 was kept at $T_0$ in all the cases. 
 Subsequently, we observed spreading 
 with an increase of 
 the liquid fraction due to condensation.

%1 
\begin{figure}[htbp]
\begin{center}
\includegraphics[scale=0.32]{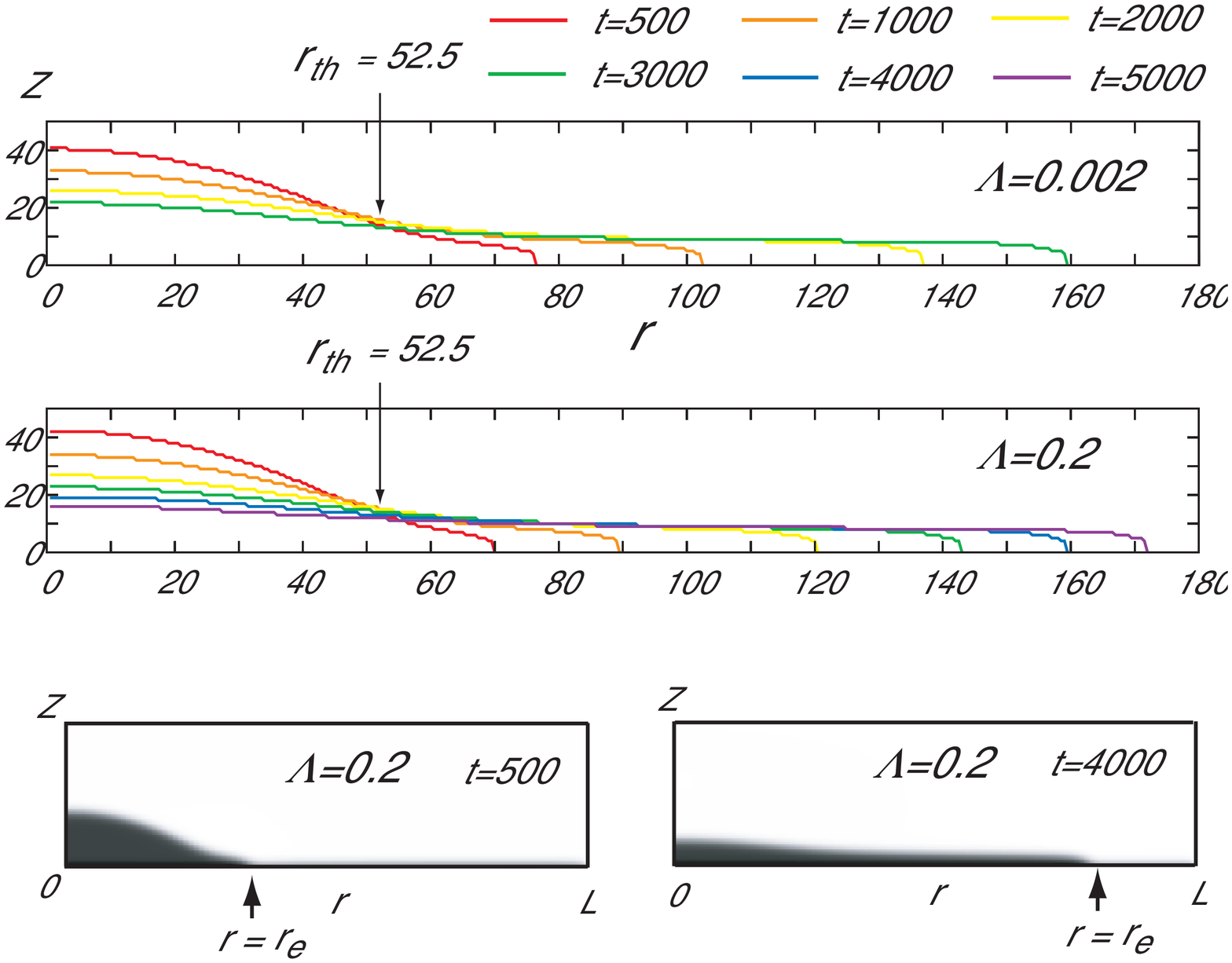}
\caption{\protect%\narrowtext 
(Color online) Shapes of an axisymmetric 
  droplet spreading on a cooler 
substrate with $T_w=0.870T_c$   
at various  times for $\Lambda=0.002$ (top) and $0.2$ (middle) 
in the $r$-$z$ plane.  
The system  temperature was initially $T_0= 0.875T_c$ at $t=0$. 
The boundary position  
between the main body of the droplet and the precursor film 
is fixed at $r=r_{{\rm th}}=52.5\ell$ for both $\Lambda$. 
The edge position $r_e(t)$ of the film increases with time 
as illustrated in the bottom plates.
}
\end{center}
\end{figure}

%%%%% Fig.2 %%%%%%%%%
\begin{figure}[htbp]
\begin{center}
\includegraphics[scale=0.5]{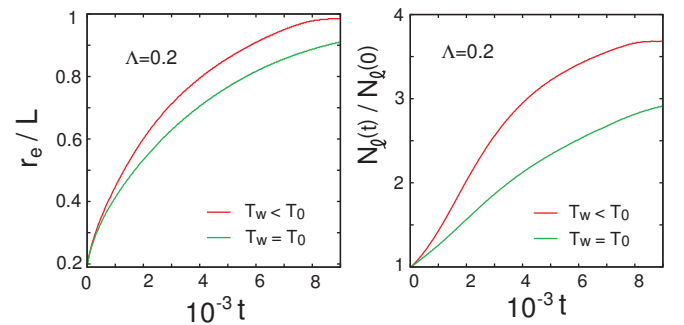}
\caption{\protect%\narrowtext 
(Color online) Time evolutions of the 
edge position $r_e(t)$ divided by $L$ 
(left)  and the particle number 
in the droplet $N_\ell(t)$ divided by $N_\ell(0)$ (right) for 
$\Lambda=0.2$. Two curves correspond to 
$T_w=0.870T_c<T_0$ (red)  and $T_w= 0.875T_c=T_0$ (green). 
The interface curve is determined by 
Eq.(4.2). The film edge  reaches  the side wall 
at $t\sim 10^4$. Condensation 
occurs faster in the cooled case 
  than in the non-cooled case.} 
\end{center}
\end{figure}

%%%%% Fig.3 %%%%%%%%%%%%%%%%%%
\begin{figure}[htbp]
\begin{center}

\includegraphics[scale=0.4]{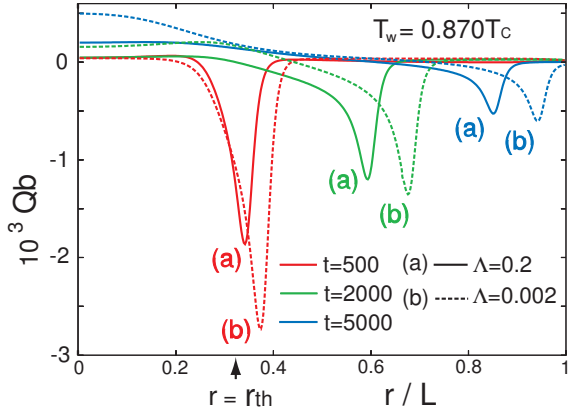}
\caption{\protect%\narrowtext 
(Color online) 
Heat flux on the substrate  $Q_{\rm b}(r,t)$ 
as a function of $r$ 
in units of $\epsilon \ell/v_0\tau_0$ 
 at various  times for $T_w=0.870T_c<T_0$ 
 with  $\Lambda=0.002$  and $0.2$.
A  negative peak at the film edge 
indicates  absorption of latent heat 
from the fluid to the solid.
At long times this absorption 
becomes  weaker and there appears a heat flow 
from the solid to the fluid 
for $r<r_{\rm th}$.}
\end{center}
\end{figure}

%%%% Fig.4 %%%%%%%%%%%%%%%%%%
\begin{figure}[htbp]
\begin{center}
\includegraphics[scale=0.45]{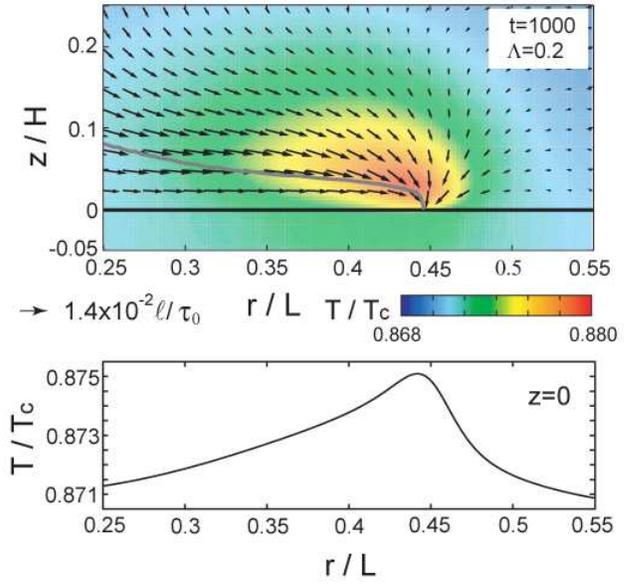}
\caption{\protect%\narrowtext 
(Color online) Temperature $T$ 
around an advancing  film edge  
at  $t=1000$, where  $T_w=0.870T_c$  and 
$\Lambda=0.2$.  
 In the top, the color represents 
the temperature according to  the color map,  and 
 the velocity field is shown by arrows  
 with its maximum being $1.4\times 10^{-2}\ell/\tau_0$ as 
 indicated below the plate.  
In the bottom,  the  substrate temperature at $z=0$ 
is plotted, which is maximum at the edge position due to 
a finite thermal conductivity of the solid. 
 }
\end{center}
\end{figure}

%5
\begin{figure}[htbp]
\begin{center}
\includegraphics[scale=0.36]{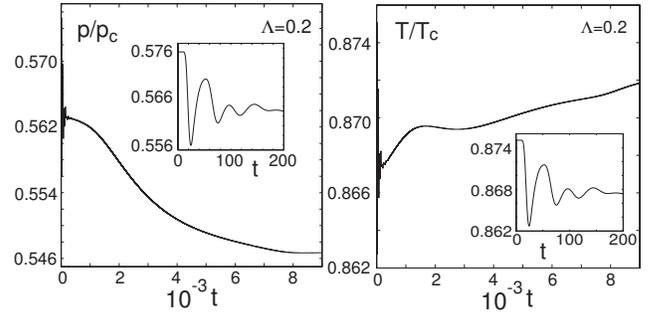}
\caption{\protect%\narrowtext 
Pressure (left) and temperature (right) vs  $t$ 
at the point $(z/H,r/L)= (0.48, 0.5)$ far from 
the substrate in gas, where   
$T_w=0.870 T_c$ and  $\Lambda=0.2$. 
Their short time behavior ($t<200$) is due to 
propagation of a 
low-pressure sound pulse 
and is  adiabatic (inset), while their long time behavior 
is due to gradual condensation.  
}
\end{center}
\end{figure}

\subsection{Evolution on long and short time scales} 

In Fig.1, the droplet  spreads 
over the substrate in the   
complete wetting condition for $\Lambda=0.002$ 
and 0.2.  
The liquid region is divided into the droplet body  
in the region  $r< r_{\rm th}$   
and the precursor film 
in  the region  $r_{\rm th}<r<r_e(t)$. 
 In our simulation,   $r_{\rm th}$ 
 is  equal to $52.5\ell=0.26L$  
 independently of time, while 
 $r_e(t)$ increased  in time. 
 The film thickness $\ell_f$ was  only weakly dependent on time  
 being  about $5 \ell$ for both $\Lambda$ (see 
 the film profiles  in Fig.6 below). 
   However, for slightly deeper cooling (say,  for $T_w=0.868T_c$) 
or for slightly larger $\Phi_1$ (say,  for  $\Phi_1 = 0.065$),  
a  new liquid region (a ring here) 
appeared on the substrate ahead of the 
precursor film.

In Fig.2, 
we show $r_e(t) $ and the particle number in the 
liquid region  $N_\ell(t)$ 
vs $t$ for $\Lambda=0.2$ 
in the cooled case with $T_w=0.870T_c$ 
and the non-cooled case with $T_w=T_0= 0.875T_c$. 
We calculate $N_{\ell}(t)$ from   
\be 
N_\ell(t) = 
2\pi    \int_0^{r_e(t)}dr r  \int_0^{z_{\rm int}(r,t)} dz 
~n({\bi r},t),
\en 
where  the interface height  is  at 
$z=z_{\rm int}(r,t)$ 
in the range  $0<r<r_e(t)$. It starts from 
the initial number  
$N_\ell (0)= 0.67 \times 10^5\ell^3/v_0$ 
and becomes a few times larger at $t\sim 10^4$. 
Here  condensation  takes place   even for 
 the non-cooled case with $T_w=T_0$.  
In these two cases, the latent heat due to 
 condensation is mostly absorbed by  the 
solid reservoir. In calculating $N_\ell(t) $ we determine 
the film height $z_{\rm int}(r,t)$ from the  relation, 
\be 
n(r, z_{\rm int},t)=  (n_\ell^0+  n_g^0)/2,
\en  
where 
$n_\ell^0= 
0.579v_0^{-1}$  and  $n_g^0= 0.123v_0^{-1}$ are the densities 
on the coexistence curve at $T=0.875T_c$. 
In our case,  the film  is so thin  and 
there is no unique  definition  of $z_{\rm int}$.

In Fig.3, we display  the heat flux on the substrate 
$Q_{\rm b}(r,t)$ for the same runs.   
 From Eq.(3.7)  it is defined 
in terms of the temperature gradient 
$T'=\p T/\p z$ as   
\be 
Q_{\rm b}(r,t)= -(\lambda_w T')_{z=-0}= 
 -(\lambda T')_{z=+0}. 
\en 
Negative peaks indicate 
absorption  of   latent heat from  
the fluid to the substrate  around  the film edge.
However, at long times ($t=5000$ in the figure) 
heat is from the solid to the fluid in the region of the 
droplet body $r<r_{\rm th}$. 
The  amplitude of $Q_{\rm b}(r,t)$ around the peak 
 is larger for $\Lambda=0.002$ 
than for $\Lambda=0.2$, obviously 
 because heat is more quickly transported for smaller $\Lambda$ 
 or for larger $\lambda_w$. 
 Also $Q_{\rm b}(r,t)$ is sensitive to $T_0-T_w$. 
For example, in the non-cooled case $T_w=T_0$, 
the minima of $Q_{\rm b}(r,t)$ 
 became  about half of those  in Fig.3 (not shown here). 
In our previous simulation \cite{Teshi}, 
  a positive peak of 
$Q_{\rm b}(r,t) $ was  found 
at  the contact line of an evaporating droplet.

 In the upper panel of Fig.4, we show the temperature 
 near the  edge at $t=1000$, where  $\Lambda=0.2$ 
 and $T_w=0.870T_c<T_0$. It  exhibits   a hot spot  
 in the gas side produced by latent heat.
 In this run, the peak height of the hot spot 
  $T_{\rm p}$  depended    on $t$ 
 as $10^{-2} (T_{\rm p}-T_w)/T_c= 1.0$, $0.7$,$0.5$, and 
 $0.4$ for $10^{-3} t =1,2,3$, and $4$. 
The  maximum of the gas velocity $v_g$ 
is $0.014 $ around the hot spot, while the 
edge speed is a few times faster as 
$d r_e/dt \sim   0.04 $. The corresponding 
Reynolds number $v_g\ell_f/\nu_0$ in the gas is very small 
$(\sim 0.07$ here). 
In the non-cooling case $T_w=T_0$  
 the peak height was reduced to  
$T_{\rm p}-T_0=0.007T_c$  and  
 $v_g$ to  $ 0.008$ at $t=10^3$. 
 In the lower panel of Fig.4, the substrate temperature 
 at $z=0$ is maximum at  the film edge. 
Such a temperature variation  
in the  solid  should be measurable 
\cite{Stephan}.

In Fig.5, we display the time evolution of 
the pressure and the temperature 
at the position $(z,r)= (0.48H, 0.5L)$ 
in the gas region far from the substrate in the case    
$T_w=0.870T_c$ and $\Lambda=0.2$. 
In the inset, their  initial deviations originate from  a 
lower-pressure sound pulse   emitted from 
the adsorption layer in Eq.(3.9). 
This acoustic process  is an example of 
 the piston effect \cite{Ferrell,Miura}. 
 In this case the thermal diffusion layer due to 
 cooling  of the substrate gives rise to 
 a smaller effect. 
The emitted pulse  traverses  the cell 
 on the acoustic time  $H/c_g\sim 50$ 
 and is  reflected at the top plate, 
where $c_g \sim 4$ is the sound velocity 
in the gas.    The deep minimum  of $T$ 
below $T_w$ and that of $p$ at $t \sim 25$  
are due to  its  first passage. 
Here the adiabatic relation 
$\delta T = (\p T/\p p)_s
\delta p$ is well satisfied for the deviations 
$\delta T=T-T_0$ and $\delta p=p-p_0$. 
The adiabatic coefficient 
$(\p T/\p p)_s$  is  equal to  $ 11 T_c/p_c$ in the gas 
and is larger than that in the 
liquid by one order of magnitude. 
On long time scales, Fig.5 shows that  
the pressure gradually decreases with progress of condensation, 
while the temperature increases for $200\ls t\ls 1500$, 
slowly decreases  for $1500 \ls t \ls 3000$, 
and again increases  for longer $t$. 
The gas temperature in the middle region 
is slightly  higher than $T_w $  by $0.002T_c$ at $t=9000$. 
We note  that the gas temperature   
is influenced by a gas flow from the droplet and behaves in a complicated 
manner. 
 
\subsection{Profiles of density, temperature, and pressure}

%6 
\begin{figure}[htbp]
\begin{center}
\includegraphics[scale=0.43]{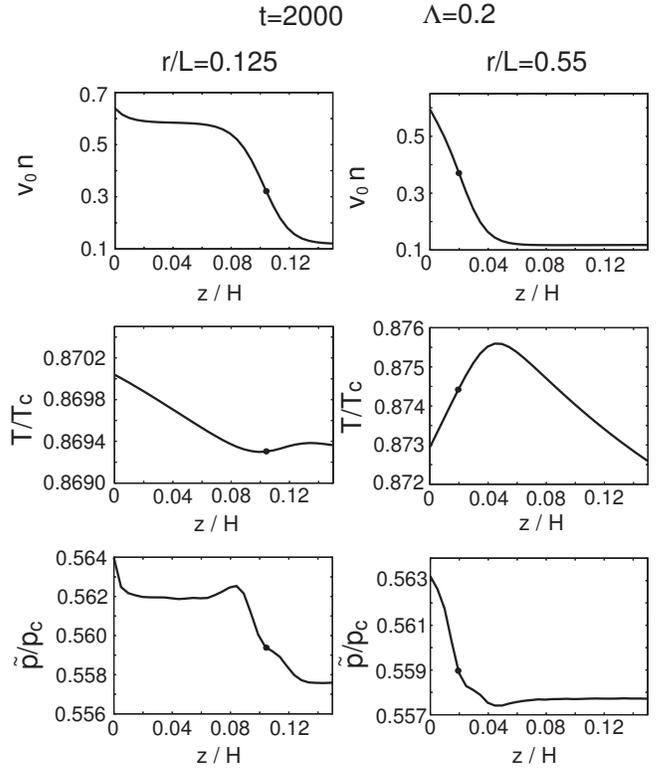}
\caption{\protect%\narrowtext 
Density (top),  temperature (middle), and 
normal pressure (bottom) 
as functions of the distance $z$ from the 
substrate   at $r/L=0.125$ (left) 
and  at $r/L=0.55$ (right), where  
 $t=2000$,  $T_w=0.870 T_c$,  and  $\Lambda=0.2$.  
The former  path passes through the droplet body, while 
the latter  through the film edge. 
The black dot $\bullet$ on each curve 
 indicates the  interface 
position  determined by Eq.(4.2).}
\end{center}
\end{figure}
%%%% Fig.7 %%%%%%%%%%%%%%%%%%
\begin{figure}[htbp]
\begin{center}
\includegraphics[scale=0.42]{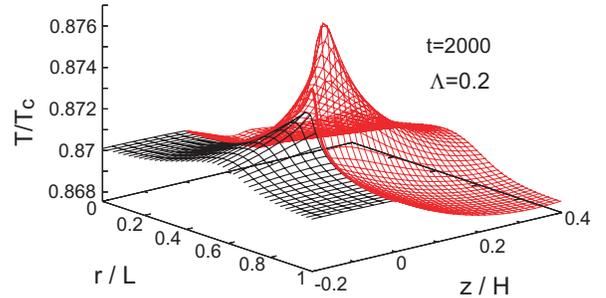}
\caption{\protect%\narrowtext 
(Color online) Temperature $T$ 
around a film edge in the fluid (red) and  
in the solid (black) in the $r$-$z$ plane 
at  $t=2000$, where  $T_w=0.870T_c$  and 
$\Lambda=0.2$. See  Fig.4  
for the hot spot at $t=1000$ in color 
in the same run.   }
\end{center}
\end{figure}

 In Fig.6, we show the profiles 
 of the density $n$, the temperature $T$, and 
 the the stress component  $\tilde p$ along 
 the density gradient at $t=2000$ 
 for $\Lambda=0.2$ and $T_w=0.870T_c$. We define 
  $\tilde p$ as 
 \be 
 {\tilde p}= \sum_{ij} \hat{\nu}_i \hat{\nu}_j 
 \Pi_{ij} =  p  - CT( n\nabla^2 n- 
 |\nabla n|^2/2), 
\en
 where $\Pi_{ij}$ is the reversible stress tensor in Eq.(2.11), 
 $p$ is the van der Waals pressure, and  
  $\hat{\bi \nu}= \{\hat{\nu}_i\}= \{\nabla_i n/ |\nabla n|\}$ 
 is the unit vector along the density gradient 
 $\nabla n$.  Thus $\tilde p$ is 
 called the normal pressure.  Obviously, ${\tilde p}\cong p$ in the bulk 
 region.    In equilibrium,  $\tilde p$   is equal to 
  the saturation pressure $ p_{\rm cx}(T)$ 
  for a planar interface \cite{OnukiV}, while it   changes 
  by the Laplace pressure difference $2\gamma/R$ 
  along $\hat{\bi \nu}$ across  
 an interface with mean curvature $1/R$. 
In nonequilibrium, we find that 
inhomogeneity of $\tilde p$ around an interface is 
much weaker than that of $p$ itself. 
The left panels for  $r=0.125L<r_{\rm th}$ in Fig.6 indicate 
weak adsorption near the wall  in Eq.(3.9),  
 a well-defined interface at $z\sim 20$, 
  and a negative temperature gradient 
  within the droplet body. 
 For this $r$,  a heat flow is from the solid to the fluid. 
  In the right panels for  $r=0.55L>r_{\rm th}$ in Fig.6, 
   $n$ decreases from  a liquid density    near the wall to a gas density 
  without a region of 
  a flat   density   and $T$ 
  exhibits a peak at the hot spot. 
  Furthermore, Fig.7 gives a bird view of 
the  temperature near the edge 
from the same run, which corresponds  
to the middle right panel in Fig.6. 
Here the temperature inhomogeneity 
in the solid can also be seen.

It is of interest how the normal pressure 
and the temperature ($\tilde p$,$T$)  at 
the interface is close to the 
coexistence line 
($p_{\rm cx}(T),T)$ in the $p$-$T$ phase diagram. 
We define   
\be 
h= \frac{T-T_0}{T_c} - \ppp{ T}{ p}{\rm cx}\frac{{\tilde p}-p_0}{T_c},
\en
where the derivative 
$({\p T}/{\p p})_{\rm cx}$  along the coexistence line is 
equal to $0.38T_c/p_c$  at $T=0.875T_c$. 
The upper panel of Fig.8 displays  $h$ around the film 
at $t=1000$,  while the lower panel of Fig.8 
gives $h$ along the surface  $z=z_{\rm int}$ 
at four times for $\Lambda=0.2$ and 0.002.
 This quantity  represents 
the distance from the coexistence 
line $p=p_{\rm cx}(T)$. In the bulk region, 
$h<0$ in stable liquid and metastable gas, 
while  $h>0$ in stable gas and metastable liquid.
We can see that $h$ 
nearly vanishes in the droplet body 
 $r<r_{\rm th}$ and increases 
in the film  $r_{\rm th}<r<r_e(t)$, but 
$h$ remains less than $10^{-2}$ even at the  edge. 
Note that the Laplace pressure 
contribution to $h$ is 
$({\p T}/{\p p})_{\rm cx}2\gamma/T_cR$, which 
is of order $0.01$ 
in the droplet body $r<r_{\rm th}$  at $t=1000$.

%%%% Fig.8 %%%%%%%%%%%%%%%%%%
\begin{figure}[htbp]
\begin{center}
\includegraphics[scale=0.31]{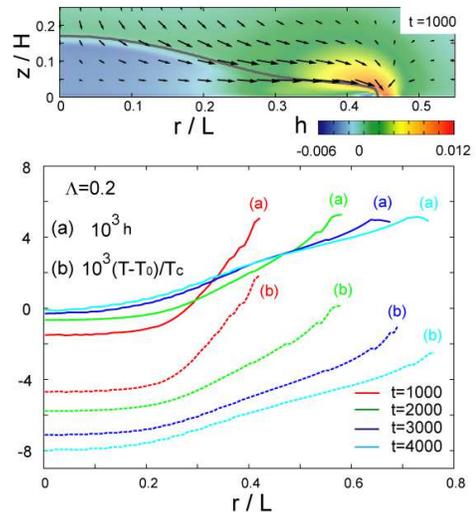}
\caption{\protect%\narrowtext 
(Color online) Top: Distance  
 from the coexistence line $h$ in Eq.(4.5) 
at $t=1000$ in color, which is negative in the droplet  body 
and is positive in the film and in the gas. 
Here  $T_w=0.870T_c$  and 
$\Lambda=0.2$. Bottom: (a) $h$ and (b) $(T-T_0)/T_c$ 
along the interface at  four  times.  
For $r<r_{\rm th}$, $|h|$ is smaller than 
$(T_0-T)/T_c$.  For $r>r_{\rm th}$, 
the distance from the coexistence line increases.  
}
\end{center}
\end{figure}

\subsection{Condensation rate and gas velocity} 

In our previous simulation\cite{Teshi},  
 evaporation of a thick liquid droplet 
mostly takes place in the vicinity of the contact line 
in the partial wetting condition.  
We here   examine the  space dependence of 
the condensation  rate of a thin fim 
in the complete wetting condition.

We introduce the number flux $J(r,t) $ from gas to liquid 
along $\hat{\bi \nu}= |\nabla n|^{-1}\nabla n$ 
through the interface, 
\be
J (r,t) =n  ({\bi v}-{\bi v}_{int})\cdot{\hat{\bi{\nu}}}, 
\en 
where  ${\bi v}_{int}$ is the interface velocity. 
If $J$ is regarded as a function 
of the coordinate along the normal direction $\hat{\bi \nu}$, 
it is continuous through 
the interface from the number conservation, 
while $ n$ and ${\bi v}\cdot\hat{\bi \nu}$ 
change discontinuously. Thus we may well determine 
$J$ on the interface.  If it is positive, it 
 represents  the local condensation rate per unit area.  
In Fig.9, we plot $J(r,t)$ vs $r/L$ 
in the region $0<r<r_e(t)$ 
at three times for $\Lambda=0.002$ and 0.2 
in the cooled case $T_w=0.870T_c$.  
We recognize that $J(r,t)$ steeply increases 
in the precursor film and is maximum at the edge. 
Moreover, it becomes   negative  in 
the body  part  $r< r_{\rm th}$ 
at $t=3000$, where evaporation occurs.  
%However, at long times $t \gs 4\times 10^4$, 
% evaporation occurs in the region 
%  $r\ls r_{\rm th}$ even for $\Lambda=0.2$, 
% as can be seen in Fig.6. 

%%%%% Fig.9%%%%%%%%%%%%%%%%%%
\begin{figure}[htbp]
\begin{center}
\includegraphics[scale=0.4]{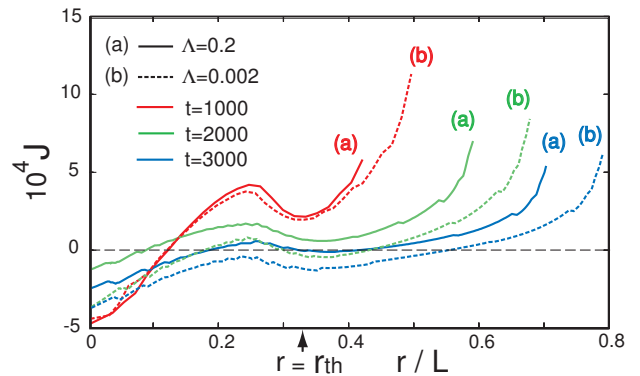}
\caption{\protect%\narrowtext 
(Color online) 
Flux $J(r,t)$  on the interface  in units of 
$\ell/v_0\tau_0$  vs $r/L$ in the region $0<r<r_e(t)$ 
 at $10^{-3} t= 1$, $2$, and $3$     
for $T_w=0.870T_c$ in the two cases of 
$\Lambda=0.002$ and 0.2.  
A precursor film 
is on the left of the arrow (see Fig.1). 
In its positive region 
it is  the  condensation rate.      In its negative region  
evaporation takes place. 
}
\end{center}
\end{figure}
%%%%% Fig.10 %%%%%%%%%%%%%%%%%%
\begin{figure}[htbp]
\begin{center}
\includegraphics[scale=0.4]{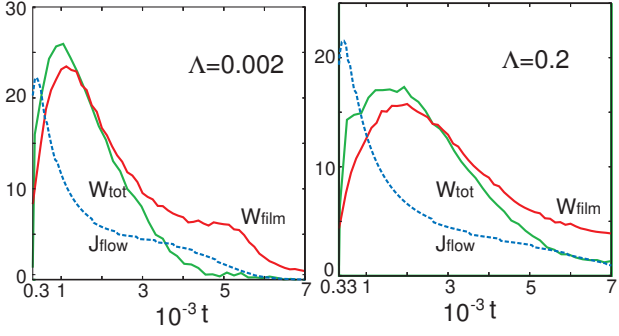}
\caption{\protect%\narrowtext 
(Color online) Total condensation rate $W_{\rm tot} (t)$ (green), 
condensation rate onto the film $W_{\rm film} (t)$ (red), and 
flow from the droplet body to the film 
$J_{\rm flow}(t)$ (blue) in units of $\ell^3/v_0\tau_0$ as functions of time. 
The time range 
is $[3\times 10^2,7\times 10^3]$ 
for $\Lambda=0.002$ (left) and 
$[3.3 \times 10^2,7\times 10^3]$ 
for $\Lambda=0.2$ (right). }
\end{center}
\end{figure}

The  total condensation rate $W_{\rm tot}(t)$ 
is the surface integral   of $J(r,t)$ on all the surface. 
The surface area in the range $[r,r+dr]$ 
is $da= 2\pi dr r/\sin\theta$, where  $\theta$ is  the angle 
between  $\hat{\bi \nu}$ and the $r$ axis. Thus, 
\be
W_{\rm tot} (t)=2\pi \int_{0}^{r_{\rm c}} dr~ {r}J(r,t)/{\sin\theta}    .  
\en 
The particle number in the liquid region  
$N_\ell(t)$ in Eq.(4.1)  
increases in time as 
\be 
\frac{d }{dt}N_\ell(t) = W_{\rm tot}(t) . 
\en 
We also define  the condensation 
rate in the film region,   
\be 
W_{\rm film}(t)
 = 2\pi \int_{r_{\rm th}}^{r_e(t)} dr~ {r}J(r,t)/{\sin\theta}  ,
\en 
where $\sin \theta\cong 1$. 
In this integral the vicinity of the edge 
gives rise to  a main  contribution. In fact,
the contribution from the 
region $r_e -16\ell <r<r_e$ 
is about $50\%$ of the total contribution 
from the region $r_{\rm th} <r<r_e$. 
 Therefore, in terms of the gas velocity  $v_g$ 
 and the gas density  $n_g$  
 around  the edge, we estimate $W_{\rm film}(t)$ as    
\be 
W_{\rm film}(t)  \sim 2\pi r_e {n}_g v_g \ell_c,
\en 
where $\ell_c$ is the width of the condensation area 
estimated to be about $30\ell$.

The  flux  from the droplet body to the film 
is given  by 
\be 
J_{\rm flow}(t)=
2\pi r_{\rm th} \int_0^{z_{\rm th}} dz~ n(r_{\rm th},z,t)  
v_r(r_{\rm th},z,t),  
\en 
where $v_r(r,z,t) = v_x x/r + v_y y/r$ is the velocity 
in the plane within the film. 
This lateral flux  
is defined at $r=r_{\rm th}$. 
More generally, we may  introduce the flux 
\be 
J_{\rm f}(r,t) = 
2\pi r \int_0^{z_{\rm th}} dz~ n(r,z,t)  v_r(r,z,t),
\en 
 for  $r\ge r_{\rm th}$  
with $z_{\rm th}$ being the film thickness $ \ell_f$. 
Then $J_{\rm flow}(t)= J_{\rm f}(r_{\rm th},t)$. 
In the presence of  condensation onto the film, 
$J_{\rm f}(r,t)$  increases  with increasing $r$. 
Its maximum  $J_{\rm f}(r_e(t),t)$  is 
estimated as $2\pi r_e {\bar n}_\ell v_\ell \ell_f$,   
where ${\bar n}_\ell$ 
is the average density in the film and 
$v_\ell$ is the average fluid velocity 
in the film.  
At $t=1000 $ (or $4000)$, 
 the ratio $J_{\rm f}(r_e(t),t)/J_{\rm f}(r_{\rm th},t) $ 
 was equal to 1.1 (or 2.4) for $\Lambda=0.2$.

In terms of  $W_{\rm film}(t)$ and $J_{\rm flow}(t)$,  
 the  particle number in the film, 
written as $N_{\rm film}(t)$, changes  in time as 
\be 
\frac{d }{dt}N_{\rm film} (t)= W_{\rm film}(t) 
+ J_{\rm flow}(t) .
\en 
Using the edge velocity $\dot{r}_e =d{r}_e/dt$, 
we also obtain  
\be
\frac{d }{dt}N_{\rm film} (t)=  2\pi r_e \dot{r}_e {\bar n}_\ell \ell_f, 
\en
since the film thickness is fixed in our case. 
In Fig.10, we plot 
$W_{\rm tot}(t)$,  $W_{\rm film}(t)$, and 
$J_{\rm flow}(t)$ vs $t$ for $\Lambda=0.002$ and 0.2. In 
an early stage ($t <1.5\times10^3$ for $\Lambda=0.002$ and 
$t <2.6\times10^3$ for $\Lambda=0.2$),  
 $W_{\rm tot}(t)$ is larger than $W_{\rm film}(t)$, 
where condensation occurs on all the interfaces. 
 Afterwards, the reverse relation 
$W_{\rm tot}(t)<
W_{\rm film}(t)$ holds, 
where evaporation occurs in the droplet body $r<r_{\rm th}$. 
We also notice $W_{\rm film}(t)>J_{\rm flow}(t)$ 
for $t \gs 1000$ for these  two $\Lambda$. 
This means that 
 the film extends  mainly 
 due to condensation near the film edge 
 except in the  early stage. 
For example,  at $t=1000$ (or $4000$),  
we have the edge velocity 
${\dot r}_e = 0.04$ (or $ 0.012$) 
and the gas velocity $v_g =0.012$ (or $ 0.0064$) 
near the edge in the case $\Lambda=0.2$. 
The fluid  velocity ${\bar v}_\ell$ 
in the film is $0.015$ (or $0.005$) at $r=r_{\rm th}$. 
These  values surely 
yield $W_{\rm film}(t)\sim J_{\rm flow}(t)$ 
 at $t=1000$ and  
$W_{\rm film}(t) \sim 3J_{\rm flow}(t)$  
at $t=4000$ in accord  with their curves in 
the right panel of Fig.10 
and are consistent with Eqs.(4.13) and (4.14).  
Thus, condensation near  the film edge 
can be the dominant mechanism of the precursor 
film growth,  as originally 
expected by Hardy \cite{Hardy,PG}.

We next estimate the gas velocity $v_g$ near the edge.  
The heat flux   is  of order 
$\lambda_\ell (T_{\rm p}-T_{w})/\ell_f$ there, 
where  $T_{\rm p}$ is 
the peak temperature  and  $\lambda_\ell$ 
is the liquid thermal conductivity. 
It balances with  
the convective latent heat flux 
$\sim n_g T_{0}\Delta s v_g$ 
in the gas, where $n_g$ is the 
gas density and $\Delta s$ is the entropy difference per 
particle.   Therefore, 
\bea 
v_g &&\sim
 \lambda_\ell(T_{\rm p}-T_{w})/
(\ell_f n_g T_{0}\Delta s) \nonumber\\
&&\sim (T_{\rm p}-T_{w}){\bar n}_\ell\nu_0/
(T_0 n_g\ell_f ) ,
\ena
where we set  $\lambda_\ell=k_B\nu_0 {\bar n}_\ell$ and 
$\Delta s(= 2.1 k_B$ here) in the second line.  
For example,  in 
 the upper  plate of 
Fig.4 at $t=1000$ 
 we have  $v_g=0.014$, while 
 the second line   of Eq.(4.15) 
 gives   0.012 with  $\ell_f/\ell\sim 5$ 
 and ${\bar n}_\ell/n_g\sim 5$.

\section{Spreading and Evaporation 
on a heated substrate}
\setcounter{equation}{0}

%11
\begin{figure}[t]
\begin{center}
\includegraphics[scale=0.34]{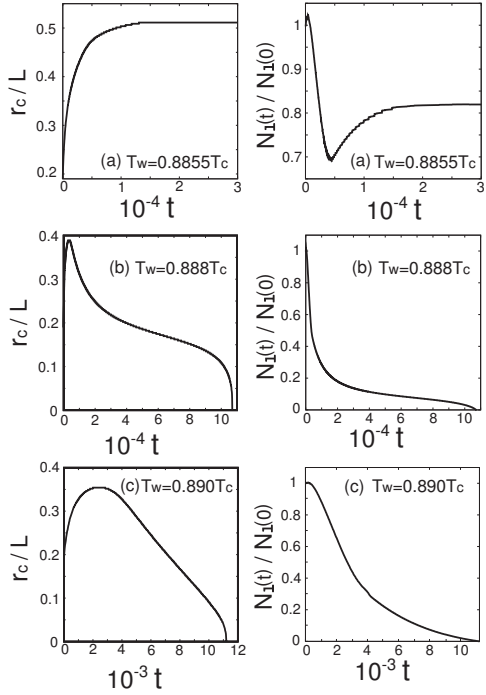}
\caption{\protect%\narrowtext 
Time evolutions of the 
edge position $r_e(t)$ divided by $L$ 
(left)  and the particle number 
in the droplet $N_\ell(t)$ divided by $N_\ell(0)$ (right) 
 for $\Lambda=0.2$. The temperature $T_w$ 
at   the  solid bottom was raised at $t=0$ from $0.875T_c$ to 
 (a) $0.8855T_c$ (top), (b) $0.888T_c$ (middle), 
and (c) $0.890T_c$ (bottom). The fluid tends to a steady 
 two-phase state in (a), where the liquid region assumes a 
pancake  thin film. The liquid  evaporates 
to vanish on a time scale of $10^5$ in (b) 
and $10^4$ in (c). 
}
\end{center}
\end{figure} 

Next, we  present simulation results of a heated liquid 
droplet in the complete wetting condition, where 
$T_w$  is increased 
above $T_0=0.875T_c$ at $t=0$ with 
 $\Lambda=0.2$. 
The  other parameter values are   the same as those 
in the previous section. The preparation method 
of a droplet is  unchanged. Then  
a precursor film  develops  
in an early stage (at least for small  $T_w-T_0$), 
 because  of the complete 
wetting condition   at $\Phi_1=0.061$ (see Eq.(3.4)).  
A new aspect is that evaporation  
 dominates over condensation  
 with increasing $T_w-T_0>0$. 
 The  experiment by 
 Gu$\acute{\rm e}$na {\it et al.}\cite{Caza}
  corresponds to this situation (see Section 1).

In Fig.11, we show  
the edge position $r_e(t)$   and the particle number 
in the liquid $N_\ell(t)$ as functions of $t$ 
for three cases 
(a) $T_w= 0.8855T_c$,  (b) $0.888T_c$, 
and (c) $0.890T_c$.  
In the weakest heating case (a) with $T_w-T_0=0.0105T_c$,   
$r_e(t)$   and  $N_\ell(t)$ tend to constants at long times, 
where  a   thin pancake-like film 
is  realized with radius $\sim 0.5L$ 
and  thickness $\sim 4\ell$ in  a steady state. 
For higher $T_w$,   
evaporation dominates over condensation 
and the liquid  region eventually disappears. 
Thus,  if $T_w-T_0$ exceeds a critical value,  
 a liquid droplet has a finite  lifetime due to evaporation  
 even in the complete wetting condition. 
From Fig.11, this lifetime is of order $10^5$  at 
$T_w-T_0=0.018T_c$  in  (b) 
and  $10^4$ at $T_w-T_0=0.020T_c$ in  (c).

%12
\begin{figure}[t]
\begin{center}
\includegraphics[scale=0.4]{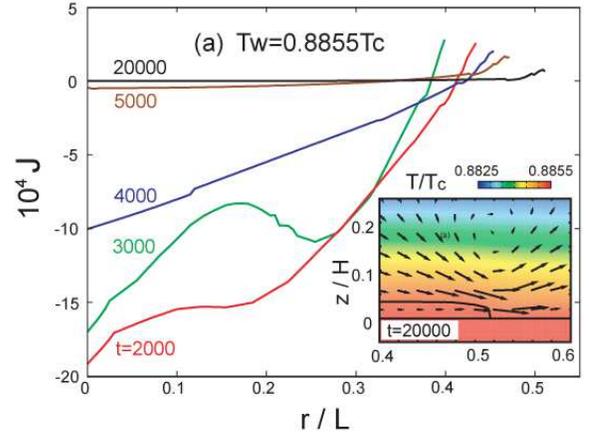}
\caption{\protect%\narrowtext 
(Color online) Mass flux $J(r,t)$ through the interface 
in Eq.(4.6) for the weakly heated case $T_w= 0.8855T_c$ 
corresponding to (a) in Fig.11. 
Evaporation takes place  
in the region $J<0$  
except close to the edge.  
Here   $J$ decreases in time. 
A steady two-phase state 
is approached at long times, 
where $J$ is  nonvanishing only near the edge. 
Inset:  temperature in color 
and velocity represented by arrows at  $t=20000$.   
}
\end{center}
\end{figure}
%13
\begin{figure}[t]
\begin{center}
\includegraphics[scale=0.37]{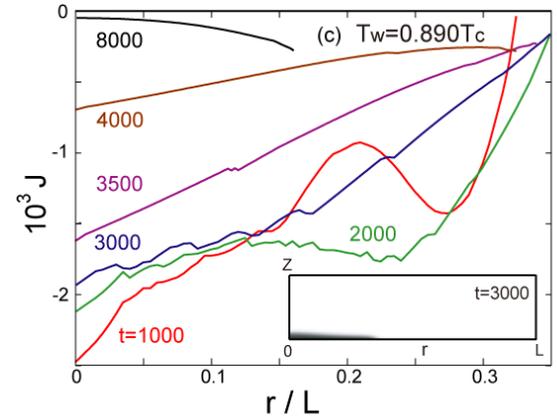}
\caption{\protect%\narrowtext 
(Color online)  Mass flux through the interface 
$J(r,t)$ in Eq.(4.6) 
for the  highest heating  case $T_w= 0.890T_c$ 
corresponding to (c) in Fig.11. Here  
evaporation takes place  over the whole  surface and 
 the liquid   disappears   
at $t=11000$. The inset displays the fim shape at $t=3000$, 
where $N_\ell(t)$ is half of the initial value.}
\end{center}
\end{figure}
%14
\begin{figure}[htbp]
\begin{center}
\includegraphics[scale=0.4]{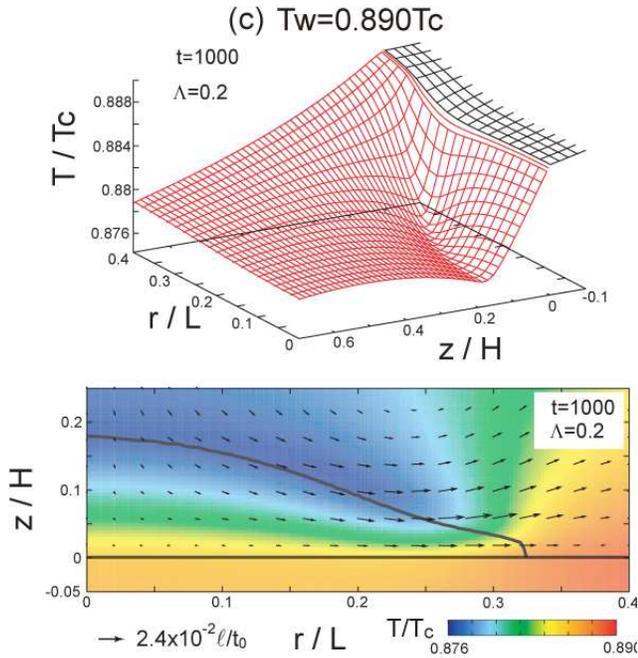}
\caption{\protect%\narrowtext 
(Color online) 
Top: Temperature $T$  in the region $0<r/L<0.4$ and 
$-0.1<z/H<0.7$ at $t=1000$ 
in  the highest heating $T_w= 0.890T_c$  
corresponding to (c) in Fig.11. 
Bottom:  temperature in color and 
velocity by arrows at the same time 
in the same run. A liquid film supports a large temperature gradient and 
  evaporation  occurs 
all over the surface, while the 
temperature above the film is nearly flat due to a gas flow. 
}
\end{center}
\end{figure}

In Fig.12, we show the mass flux 
 through the interface $J(r,t)$ defined in Eq.(4.6) 
in  the weakly heated  case (a) $T_w= 0.8855T_c$.
Its negativity implies evaporation. 
 In the region far from the edge,  
 evaporation is marked 
in transient states ($t\ls 4000$), 
 but it tends to vanish at long times. 
We can also see the region 
of positive $J$ with width of order 10 
near  the edge ($r_e-10<r<r_e$), 
where the film is still flat 
and the angle $\theta$ in Eq.(4.7) 
is nearly $ \pi/2$. In Fig.12, however,  
we do not show $J$ just at the edge 
($r\cong r_e(t)$ and $0<z<\ell_f$),  where  
 $\theta$ changes from $\pi/2$ to 
 zero in  the $z$ direction  
and evaporation occurs ($J<0$). 
As a balance of  condensation and evaporation 
in these two regions, the total condensation rate 
$W_{\rm tot}$ in Eq.(4.7) tends  to 
vanish at long times, while there is no velocity 
field in the region  $r<r_e-10$. 
 In the inset of Fig.12, 
the velocity field  around the edge is displayed 
at $t=20000$,  where the maximum 
gas velocity is  $v_g= 1.1\times 10^{-3}\ell/\tau_0$.  

%In the final steady state at $t=20000$, 
%evaporation and condensation 
%can occur only near  the film edge. 

In Fig.13, we show  $J(r,t)$ 
  at several times in  the highest heating    
  case (c) $T_w= 0.890T_c$.  
In  the whole  surface, $J$ is negative 
and evaporation ocuurs. 
For $t\ls 40000$ evaporation is strongest at the fim center. 
At long times ($t=8000$ here), it becomes 
weakest at the film center. 
 Figure 14 is produced by 
 the same run. It  gives  a bird view of 
 the temperature and a snapshot 
 of the velocity field 
 in the vicinity of the film edge 
 at $t=1000$.  
We can see a steep temperature gradient 
within the film, which is much larger than in the gas, 
leading to a strong heat flux from the solid to 
the film. In this manner,  evaporation is induced  
all over the surface and is strongest 
at the film center in the early stage. 
It is  remarkable that 
the temperature gradient nearly vanishes 
in the gas region above  the film 
away  from the edge, where heat is transported 
by a gas flow.  
We can also see a significant 
 temperature inhomogeneity in the solid part 
 in contact with the film.

\section{Summary and remarks}

For one-component fluids we 
 have  examined  spreading  of a small droplet 
on a  smooth substrate  
in the complete wetting condition 
in the axisymmetric geometry.
In the dynamic van der Waals 
theory \cite{OnukiPRL,OnukiV}, 
we have integrated the entropy equation in Eq.(2.14)  
together with the continuity and momentum equations. 
This method  may remove artificial 
 flows around an interface \cite{para}.  
In our phase field scheme, 
we need not introduce any surface 
boundary conditions. 
The condensation rate on the 
interface is a result  and  not a 
   prerequisite of the calculation. 
We have also assumed that 
the substrate wall has a finite thickness $H_w$ 
and the solid temperature obeys the thermal diffusion equation, 
whereas an isothermal substrate is usually assumed in 
the literature. 
The temperature  $T_w$ at the solid bottom $z=-H_w$ 
is a new control parameter in our simulation. 
 Cooling (Heating) the fluid is realized  by 
setting $T_w$ lower (higher) 
than the  initial fluid temperature $T_0$. 
We give salient results in our simulation.   

(i) In the cooled and non-cooled 
cases with $T_w \le T_0$, a  precursor film 
with a constant thickness has appeared 
 ahead of the droplet body. 
Here the liquid volume  
has increased  in time due to 
condensation  on the film surface. 
In an very early stage, the piston  effect  comes  
into play due to sound  propagation \cite{Ferrell,Miura}.
At long times, 
the condensation rate has become 
localized near the film edge and 
the film has expanded 
dominantly  due to condensation. 
 As a result, a hot spot has appeared  
near the film edge 
 due to the latent heat released.

(ii) At a critical value of 
$T_w$ slightly higher  than $T_0$, 
 we have realized 
a steady-state  thin liquid  film, where 
condensation and evaporation are 
localized and balanced at the edge. 
For higher   $T_w$, evaporation  has dominated   
and the liquid region has disappeared  eventually.     
This lifetime decreases with increasing  $T_w-T_0$. 
For a thin film,  evaporation has appeared   
all over the film surface upon heating. 
 In our previous simulation for 
one-component fluids \cite{Teshi} 
 evaporation of a thick 
droplet was mostly localized  near the contact line 
  in  the partial wetting condition.

We give some critical remarks.   
(1) If the mesh length $\Delta x=\ell/2$ 
is a few $\rm \AA$, our system length  is on the order of 
several ten manometers and the particle number treated 
is of order $10^6$ (see Sec.IIIC). Our continuum 
description should be imprecise on the angstrom scale.  
Thus examination  of our results by large-scale 
molecular dynamics simulations should  be informative. 
We should also investigate  
how our numerical results can be used or modified  
for much larger  droplet sizes.  
(2) In future work, we shoud examine 
 the  role of the long-range van der Waals interaction 
 in the wetting  dynamics. As is well-known,  
it   crucially 
influences  the film thickness  \cite{PG}. 
(3)  We should also include  the slip effect  
  at the contact line in our scheme \cite{slip,Qian}.  
(4) We should study the two-phase 
hydrodynamics  in fluid mixtures, where  
a Marangoni flow decisively governs 
the dynamics even at small solute concentrations 
 \cite{Str,Larson,maran}.

\begin{acknowledgments}
This work was supported by Grants-in-Aid 
for scientific research 
on Priority Area ``Soft Matter Physics" 
and  the Global COE program 
``The Next Generation of Physics, Spun from Universality and Emergence" 
of Kyoto University 
 from the Ministry of Education, 
Culture, Sports, Science and Technology of Japan. 
\end{acknowledgments}


\begin{references}
\bibitem{gravity} If we set 
$e_{\rm T}=\hat{e}+ \rho {\bi v}^2/2$ 
not including  the potential $U$, 
 the term $-\rho {\bi v}\cdot \nabla U$ 
appears in the right hand side of Eq. (2.13) \cite{OnukiV}.
\end{references}
\end{document}